\pdfoutput=1
\documentclass[acmlarge]{acmart}

\usepackage[table,xcdraw]{}

\usepackage{geometry}
\usepackage[english]{babel}     
\usepackage{caption}
\usepackage{makecell}
\usepackage{hyperref}
\usepackage{float}
\floatstyle{plaintop}
\restylefloat{table}

\setcellgapes{3pt}

\setcellgapes{3pt}

\usepackage{stfloats}

\usepackage{multirow}
\AtBeginDocument{%
  }

\setcopyright{acmcopyright}
\copyrightyear{2023}
\acmYear{2023}
\acmDOI{XXXXXXX.XXXXXXX}

\acmJournal{POMACS}
\acmVolume{37}
\acmNumber{4}
\acmArticle{111}
\acmMonth{2}

\begin{document}

\title{Automated Cyber Defence: A Review}

\author{Sanyam Vyas}
\email{vyass3@cardiff.ac.uk}
\orcid{0000-0002-5603-6868}
\affiliation{%
  \institution{Cardiff University}
  \city{Cardiff}
  \state{Wales}
  \country{United Kingdom}
  \postcode{CF24 4AG}
}

\author{John Hannay}
\email{hannayj1@cardiff.ac.uk}
\orcid{0000-0002-7853-0824}
\affiliation{%
  \institution{Cardiff University}
  \city{Cardiff}
  \state{Wales}
  \country{United Kingdom}
  \postcode{CF24 4AG}
}

\author{Andrew Bolton}
\email{boltona2@cardiff.ac.uk}
\orcid{}
\affiliation{%
  \institution{Cardiff University}
  \city{Cardiff}
  \state{Wales}
  \country{United Kingdom}
  \postcode{CF24 4AG}
}

\author{Pete Burnap}
\email{burnapp@cardiff.ac.uk}
\orcid{0000-0003-0396-633X}
\affiliation{%
  \institution{Cardiff University}
  \city{Cardiff}
  \state{Wales}
  \country{United Kingdom}
  \postcode{CF24 4AG}
}

\renewcommand{\shortauthors}{Vyas et al.}

\begin{abstract}
Within recent times, cybercriminals have curated a variety of organised and resolute cyber attacks within a range of cyber systems, leading to consequential ramifications to private and governmental institutions. Current security-based automation and orchestrations focus on automating fixed purpose and hard-coded solutions, which are easily surpassed by modern-day cyber attacks. Research within Automated Cyber Defence will allow the development and enabling intelligence response by autonomously defending networked systems through sequential decision-making agents. This article comprehensively elaborates the developments within Automated Cyber Defence through a requirement analysis divided into two sub-areas, namely, automated defence and attack agents and Autonomous Cyber Operation (ACO) Gyms. The requirement analysis allows the comparison of automated agents and highlights the importance of ACO Gyms for their continual development. The requirement analysis is also used to critique ACO Gyms with an overall aim to develop them for deploying automated agents within real-world networked systems. Relevant future challenges were addressed from the overall analysis to accelerate development within the area of Automated Cyber Defence.   
\end{abstract}

\ccsdesc[100]{Computer systems organization~Artificial Intelligence; Reinforcement Learning}
\ccsdesc[100]{Security and privacy ~ Network Security; Malware Mitigation}

\keywords{automated cyber defence, reinforcement learning, intrusion response, network security}


\maketitle

\section{Introduction}
Individuals, organisations and governments across the world are facing an exponentially increasing digital involvement within their daily activities and operations. While this has efficiently connected the world together in terms of information awareness and communication, all entities mentioned above are continuously facing cyber attacks from a range of attackers such as opportunists, criminals and hostile states. The exponential growth of such Information Technology (IT) and Operational Technology (OT) devices within homes and industry, along with an increasing skills shortage has led to cybersecurity infrastructures and organisations being overwhelmed in lieu of the increasing amounts of cyber attacks that have occurred. While there is a desperate requirement of skilled cybersecurity practitioners, the level of recent novel and automated cyber attacks surpass the ability of humans manually defending against them \cite{bridges2022testing}. Therefore, there is now a desperate requirement for automated defence solutions \cite{kinyua-et-al-ai-cysec, mahaini-et-al-automation} to be implemented within IT and OT infrastructures in order to manage such threats against such systems. Consequently, while there have been automated defence solutions within literature, their inability to defend against recent cyber attacks require more research to be conducted to limit the monetary and intellectual property based damages. \\
\\
As a solution to this problem, we introduce Automated Cyber Defence, an area that focuses on automated decision-making agents for networked systems to mitigate highly complex cyber attacks. This paper defines ACD and analyses literature within different divisions of Automated Cyber Defence. The analysis is conducted through an overall requirement analysis with the vision of focusing on the real-world deployment of such automated decision-making agents within networked systems. Overall, very few publications have highlighted the requirement of automated decision-making agents for defending against cyber-attacks within networked systems. Recent publications include include \cite{huang-et-al-rl-crm} which provides a detailed review on Reinforcement Learning (RL) solutions for moving target defence, cyber defence and honeypots. The publication also provides a detailed development of RL solutions within cybersecurity through optimal control-theoretic principles and latest AI developments. However, the review does not focus on terrains for the rapid development of such automated decision-making agents for network defence and attack. Wang et al \cite{wang2022research} also focus on the development of RL solutions for network defence and attack, along with addressing future challenges similar to \cite{huang-et-al-rl-crm}. However, the paper did not exhaustively analyse the terrains on which such agents could be developed, which is an imperative part of Automated Cyber Defence. Authors from \cite{buettner-et-al-cy-def-review} provide a review on Machine Learning (ML) solutions within cybersecurity, specifically focusing on data sets that accelerate research within intrusion detection systems. While this implementation forms a part of automation approaches within cybersecurity, intrusion detection approaches do not apply to the Automated Cyber Defence term defined in this paper as they do not involve an automated cyber response. Burke et al \cite{burke-et-al-turing} provide an in-depth review on the type of potential projects within Active Cyber Defence (AcCD), some of which also apply to the term Automated Cyber Defence mentioned in this paper. However, the report does not focus on the ACO gyms for development, and nor does it provide a detailed comparison of automated decision-making algorithms used for automated network attack and defence.\\
\\
The rest of this article includes section \ref{sec: key-definitions} that firstly defines the different terms used frequently in this paper. Section \ref{fig: res-methodology} then addresses the methodology utilised to find relevant ACD publications. Subsequently, section \ref{sec:acd} then elaborates the curated terminology of Automated Cyber Defence and its differentiation from similar terminologies used within recent literature. The section then provides the importance of the area in National Strategies. Lastly, the section provides a comprehensive Requirement Analysis that will be used to evaluate the selected publications recognised to be as part of Automated Cyber Defence. Section \ref{sec: acd-custom-aco-gym} elaborates and critiques on the automated defence and attack (blue and red) agents in custom ACO Gyms through the Requirement Analysis in Section \ref{sec:acd}. Section \ref{sec: aco-gyms} elaborates an exhaustive list open-source and closed-source ACO Gyms and assesses them using the Requirement Analysis in Section \ref{sec:acd}. Section \ref{sec: acd-custom-aco-gym} elaborates a list of published automated agents within ACO Gyms and evaluates them using the Requirement Analysis in section \ref{sec:acd}. Section \ref{sec: challenges} provides a discussion identifying the challenges and gaps within Automated Cyber Defence literature using the assessments conducted in the previous sections. Lastly, section \ref{sec: conc} concludes the article by summarising the area of Automated Cyber Defence. The contributions of this paper include:

\begin{itemize}
\item Complete definition of the term Automated Cyber Defence and distinguishes it's research as compared to other related terms.
\item Development of a Requirement Analysis for the defined field of Automated Cyber Defence, highlighting the requirements of two important areas within Automated Cyber Defence, namely, the development criteria of Automated Blue and Red Agents, and the development criteria of ACO Gyms to facilitate Automated Cyber Defence capabilities.
\item Assessment of the publications within Automated Cyber Defence literature through the requirement analysis.
\item Identification of novel and realistic challenges within the literature to highlight future novel research directions. 
\end{itemize}

\section{Key Definitions}
\label{sec: key-definitions}

This article comprises of several technical terminologies that are commonly used within the fields of cybersecurity and artificial intelligence. This section will define the key terminologies used within this document.\\
\\
\textbf{Automated Red Teaming:} Red Teaming is a technique used within military and industry operations to uncover networked system vulnerabilities or to find exploitable gaps in operational concepts, with the overall goal of reducing surprises, improving and ensuring the robustness of the networked system. \cite{red-teaming-def}. In the context of this paper, automated red teaming refers to an autonomous agent possessing a set of operations (to uncover vulnerabilities and exploitative within the networked system) as their action space. The overall aim of automated red teaming is to ensure the robustness of the automated blue team agent (definition elaborated below) in terms of defending the system against known vulnerabilities and exploits.\\
\textbf{Automated Blue Teaming:} Blue teaming is a technique responsible for defending a networked systems by maintaining its security posture against a group of mock attackers that aim to exploit gaps and vulnerabilities of the networked system. Typically the Blue Team and its supporters must defend against real or simulated attacks 1) over a significant period of time and 2) in a representative operational context (e.g., as part of an operational exercise)\footnote{https://csrc.nist.gov/glossary/term/blue\_team}. In the context of this paper, Automated Blue Teaming refers to an autonomous agent possessing a set of operations as their action space to destroy malicious processes from entering the networked system through its nodes/endpoints. \\
\textbf{Autonomous Cyber Operations Gym:} Autonomous Cyber Operations (ACO) is concerned with the defence of computer systems and networks through autonomous decision-making and action. It is particularly required where the deployment of security experts to cover every network and location is becoming increasingly untenable, and where systems cannot be reliably accessed by human defenders, either due to unreliable communication channels or adversary action. ACO Gyms are networked system environments that facilitate the use of autonomous red and blue teaming agents in order to further strengthen the networked systems of the future from ever-evolving cyber attacks \cite{standen2021cyborg}. ACO Gyms aim  to address and reduce the ‘reality gap’ of potential networked systems, used in \cite{tan-et-al-aco} by combining learning on simulations with testing in a real environment.\\
\textbf{Sequential response:} Sequential response, or sequential decision-making refers to algorithms that take the dynamics of the world into consideration, thus delaying segments of the problem until it is solved \cite{frankish_ramsey_2014-sequential-response}. It is a fundamental task faced by any intelligent agent in an extended interaction with its environment which demands a set of decisions that are concerned with short and long-term decisions in order to reach a state that acts as an overall target within the environment.\cite{littman-sequential-response}. In the context of this paper, sequential decision-making algorithms are considered in this paper as Automated Blue and Red Teaming agents due to the complexity of the network that requires navigation before a target action is taken by the automated agent (e.g. launching an exploit in a host within a different subnet).\\
\textbf{Single-step response:} Single-step response algorithm refers to decision-making actions that only focus on the short-term outcomes. For example, in temporal context, the algorithm at time \textit{t(n)} will perform calculations solely for a solution at time \textit{t(n + 1)}. \\
\textbf{Simulated Network:} A Simulated Network is an ACO Gym (or a part of the ACO Gym's training-testing strategy) that is designed as a finite state machine. The creation is usually completed in the form of code that includes objects that correspond to the components, agents and actions within the simulated network.  \cite{farland-MolinaMarkham-et-al}\\
\textbf{Emulated Network:} An Emulated Network is an ACO Gym (or a part of the ACO Gym's training-testing strategy) that is designed through a group of virtual machines, which are used to create a computer networked system \cite{farland-MolinaMarkham-et-al}.

\section{Review Methodology}
\label{sec: review-methodology}

A methodology inspired by \cite{Kitchenham-et-al-search-strategy} was implemented to find all relevant articles for this review. In order to curate the overall ACD definition and the research questions for this article, papers from national and international government institutions and private organisations (mentioned in section \ref{subsec: strategy}) were utilised. These papers addressed the need for autonomous response solutions in networked systems within a variety of different areas. This allowed us to categorise areas where autonomous response could be utilised within the existing areas of Automated Cyber Defence terminology, specifically, Automated Red and Blue Teaming.  
\subsection{Research Questions}
In order to utilise the ideas suggested for Automated Cyber Defence, research questions were created in order to identify a search strategy. These also allowed us to find relevant literature for this article. The third research question was identified by the search strategy (elaborated in the next subsection), leading to the updating of the overall search strategy.

The research questions (RQs) included:
\begin{itemize}
\item \textbf{RQ1}: What is Automated Cyber Defence?
\item \textbf{RQ2}:  What are the most suitable algorithmic approaches that have been used within the Automated Cyber Defence terminology defined through \textbf{RQ1}? 
\item \textbf{RQ3}: What are the best possible environments in which the most suitable algorithmic approaches could be developed?
\end{itemize}

\subsection{Search Terminology Strategy}
After identifying all research questions overall, the next step involved searching for relevant primary studies. Popular digital libraries including IEEE, ACM Digital Library, Springer and Science Direct are utilised along with Google Scholar in order to not overlook significant relevant work. A list of strings grouped within 3 overall themes of Automated Cyber Defence were collectively identified (shown in Table \ref{tab:overarching-themes}). The strings from all different overall themes are then grouped together in 3 different groups of permutation combinations as an aim to identify publications in digital libraries that:
\begin{itemize}
\item \textbf{i}:  allow us to explore and rank the performance of identified algorithmic families in Automated Blue Teaming (\textbf{RQ1, RQ2}).
\item \textbf{ii}: allow us to explore and rank the performance of identified algorithmic families in Automated Red Teaming  (\textbf{RG1, RQ2}).
\item \textbf{iii}: allow us to discover the best possible environments in which the most suitable algorithms could be developed, trained and tested (\textbf{RQ1, RQ3}).
\end{itemize}

\subsection{Overall Relevant Content Extraction}

\begin{figure}[]
\centering
\includegraphics[width= 10cm]{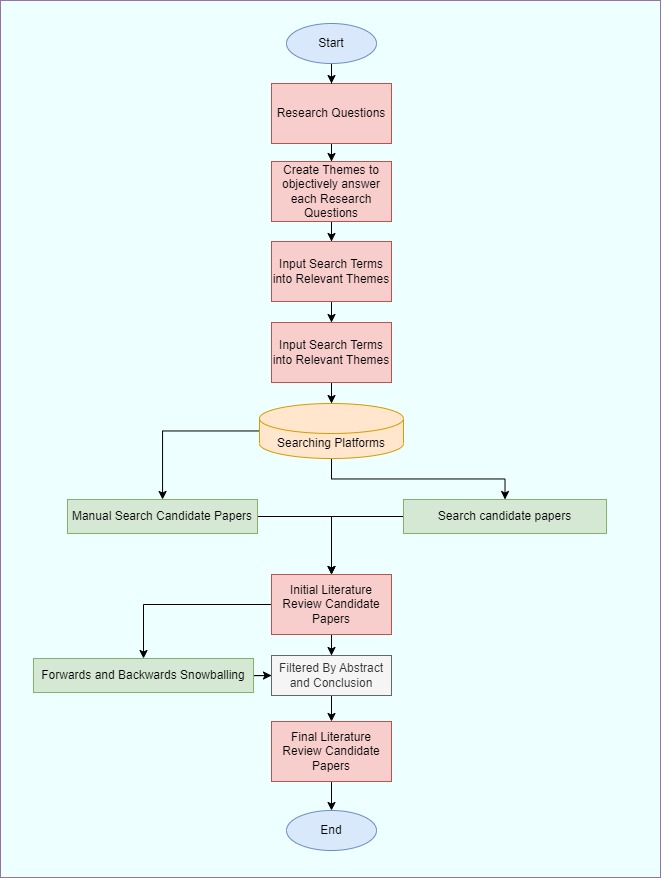}
\caption{Research Methodology}
\label{fig: res-methodology}
\end{figure}

Due to the area of Automated Cyber Defence gaining popularity only recently, backward snowballing \cite{backward-snowballing-keele} for several searches was conducted in order to find publications identified as Automated Cyber Defence that were not listed in the search strategy. For example, with areas such as "autonomous cyber operations gym" being a recently created terminology within this area, backward snowballing aided us to identify other popular publications (and implementations found in code repositories) that were created before this term was officially introduced. In addition, a manual search was conducted to identify the latest Automated Cyber Defence related papers (along with papers highlighting further potential areas within the domain) that cited the publications identified through the search strategy. All papers selected were passed through another screening process based on the papers abstract in order to align the scope of papers based on the definition overall definition of Automated Cyber Defence. Lastly, the remaining papers were then fully read and analyzed for further screening. Figure \ref{fig: res-methodology} suggests the overall steps included within this search methodology.

\begin{table}[h]
\centering

\resizebox{\textwidth}{!}{%
\begin{tabular}{llllllllll}
\hline
\multicolumn{4}{c}{\textbf{\begin{tabular}[c]{@{}c@{}}a. Algorithmic \\ Approaches \end{tabular}}}                  & \multicolumn{2}{c}{\textbf{\begin{tabular}[c]{@{}c@{}}b. Automated \\ Blue-Teaming\end{tabular}}}              & \multicolumn{4}{c}{\textbf{\begin{tabular}[c]{@{}c@{}}c. Automated \\ Red-Teaming\end{tabular}}}            \\ \hline \hline
\multicolumn{4}{l}{- "artificial intelligence"}                                                                                  & \multicolumn{2}{l}{- "autonomous cyber operations gym"}                                                                       & \multicolumn{4}{l}{- "malware"}                                                                               \\ 
\multicolumn{4}{l}{\begin{tabular}[c]{@{}l@{}}- "machine learning" \\ OR "deep learning"\end{tabular}}                           & \multicolumn{2}{l}{- "process killing"}                                                                          & \multicolumn{4}{l}{- "process"}                                                                               \\ 
\multicolumn{4}{l}{- "open ai" AND "gym"}                                                                                        & \multicolumn{2}{l}{- "cyber defence"   OR "cyber defense"}                                                       & \multicolumn{4}{l}{- "penetration" AND "testing"}                                                             \\
\multicolumn{4}{l}{- "reinforcement   learning"}                                                                                 & \multicolumn{2}{l}{- "malware"}                                                                                  & \multicolumn{4}{l}{- "offensive cybersecurity"}                                                              \\ 
\multicolumn{4}{l}{- "game theory"}                                                                                            & \multicolumn{2}{l}{- "deception"}                                                                                & \multicolumn{4}{l}{- "autonomous malware"}                                                                    \\ 
\multicolumn{4}{l}{- "generative   modelling"}                                                                                   & \multicolumn{2}{l}{-"response"}                                                                                 & \multicolumn{4}{l}{- "privilege escalation"}                                                                  \\ 
\multicolumn{4}{l}{\begin{tabular}[c]{@{}l@{}}- "automated" OR "automatic" \\ OR   "autonomous" OR \\ "automation"\end{tabular}} & \multicolumn{2}{l}{- "wargaming" OR   "war-gaming"}                                                              & \multicolumn{4}{l}{- "adversary emulation"}                                                                   \\ 
\multicolumn{4}{l}{- "response"}                                                                                & \multicolumn{2}{l}{- "cyber resilience"}                                                                         & \multicolumn{4}{l}{\begin{tabular}[c]{@{}l@{}}- "wargaming" OR \\ "war-gaming"\end{tabular}}                  \\ 
\multicolumn{4}{l}{}                                                                                                           & \multicolumn{2}{l}{\begin{tabular}[c]{@{}l@{}}- "advanced persistent threats" \\ OR "APT"\end{tabular}}          & \multicolumn{4}{l}{\begin{tabular}[c]{@{}l@{}}- "red team" OR "red teaming" \\ OR "red-teaming"\end{tabular}} \\
\multicolumn{4}{l}{}                                                                                                           & \multicolumn{2}{l}{\begin{tabular}[c]{@{}l@{}}- "blue team" OR "blue-teaming" \\ OR "blue teaming"\end{tabular}} & \multicolumn{4}{l}{- "reconnaissance"}                                                                        \\  
\multicolumn{4}{l}{}                                                                                                           & \multicolumn{2}{l}{\multirow{3}{*}{- "cyber threat intelligence"}}                                               & \multicolumn{4}{l}{- "autonomous cyber operations gym"}                                                                      \\  
\multicolumn{4}{l}{}                                                                                                           & \multicolumn{2}{l}{}                                                                                           & \multicolumn{4}{l}{\begin{tabular}[c]{@{}l@{}}- "cyber defence" OR \\ "cyber defense"\end{tabular}}           \\ 
\multicolumn{4}{l}{}                                                                                                           & \multicolumn{2}{l}{}                                                                                           & \multicolumn{4}{l}{- "deception"}                                                                             \\ \hline
\end{tabular}%
}
\caption{Suggests the overarching themes for search terminology. \textbf{a.} includes algorithms and terminologies that incorporate an automated response (a requirement for Automated Cyber Defence agents). \textbf{b.} includes all terminologies that are a part of the defined Automated Blue Teaming terminology. \textbf{c. } includes all terminologies that are a part of the defined Automated Red Teaming terminology. }
\label{tab:overarching-themes}
\end{table}

\section{Automated Cyber Defence}
\label{sec:acd}
Automated Cyber Defence (ACD) is a topic that has recently been mentioned within a few publications and news articles over the last decade, in light of the increasing cyber attacks that have occurred over the last few years. In order to define this term, a brief review was completed. \\
\\
Rege et al \cite{rege-et-al-acd} provided a high-level description of ACD algorithms as a decision-making system with expert-level ability inspired by how humans reason and learn, citing a publication \cite{benjamin-et-al-acd} producing an automated blue agent within a custom networked system. Ko et al \cite{ko-et-al} provided a terminology for ACD when elaborating the purpose of the Defense Advanced Research Projects Agency (DARPA) grand challenge \footnote{https://www.darpa.mil/program/cyber-grand-challenge}, where it described ACD as systems that are able to self-discover, prove, and correct software vulnerabilities in real-time – without human intervention. In 2016, Baah et al \cite{baah-et-al} provided a generalised overview of an ACD system. The paper described ACD as a response that begins with detection of an ongoing attack or an existing vulnerability in the network. The paper highlighted that speed and accuracy of detection is important in order to take action to mitigate threats before they can do damage to network assets or disrupt missions. It also illuminates a solution of machine learning analytics that can distinguish between suspicious and benign network activity, and automated fuzzing techniques that can discover previously unknown vulnerabilities in software. Benjamin et al \cite{benjamin-et-al-acd} define the ACD term through their project called Cognitive Support for Intelligent Survivability Management (CSISM), where the authors implement an automated cyber defence decision-making mechanism with expert level ability. The ACD system interprets alerts and observation, and then takes defensive actions to ensure the survivability of the computing capability of the network. The authors realise that producing such an expert-level response in real-time with uncertain and incomplete information is a difficult target. However, they realise that there is a stepping-stone between the development of automated reasoning and learning through the use of cognitive architectures for cyber defence operations. \\
\\
Burke et al \cite{burke-et-al-turing} from the Alan Turing Institute introduced a research initiative focusing on Active Cyber Defence (AcCD) through a white paper, which focuses on seeking increased automation within an enterprise to bolster network defenders and cybersecurity.  Note, it is important to address the difference between the term AcCD and ACD is the inclusion of Automated Security Planner within AcCD, while ACD strictly focuses automated red and blue-teaming, primarily for the overall development of automated blue teaming agents. Overall, the paper explains that intelligent automation is essential to enable system defenders to manage the risk posed by highly automated future threats and attack, and defend the systems at cyber-relevant national scale. The white paper also elaborated the need for automated red and blue teaming. However, it only provided high-level information on the research directions within the area. The use of Artificial Intelligence has been suggested within such systems as a way to intelligently understand the terrain (i.e., networked system) for detecting and responding to complex cyber-attacks with minimal errors. \\
\\
Applebaum et al \cite{applebaum2022bridging} highlight the terminology, Autonomous Cyber Defense, within their paper that utilises \textit{autonomous cyber defence} (ACD) agents based on tabular Q-learning. The terminology suggested that autonomous cyber defence is the leveraging of ML techniques to train an agent that is able to autonomously defend a system, minimising self-damage from responses that use noisy sensor data. While this definition aligns to our definition of ACD, the definition is very brief and must be expanded to understand the whole area of ACD that includes a parallel development of Autonomous Cyber Operations Gyms along with automated red and blue team agents. In addition, the defined terminology does not incorporate the role of automated red team agents, which is addressed in this paper as an imperative part of ACD.
\\
\\The definition of Autonomous Cyber Operations (ACO) will also need to be addressed relative to ACD in order to clarify specific research directions within ACD as compared to ACO. \cite{standen2021cyborg} defines ACO as the parallel development of automated red (attacker) and automated blue (defender) agents within a networked system that combat one another in a game-playing scenario. ACD differs from ACO through ACD's focus being on the overall development of automated blue agent through the automated red agents being particularly designed as an automated penetration testing agent that also facilitates adversarial training. The development of ACO Gyms in the lens of ACD also differs to the development of AI ACO Gyms in that they must be designed to specifically for the development of automated blue teaming agents. \\
\\
When compiling all literature's mentioned above, we define \textbf{Automated Cyber Defence} as a terminology focusing on the automated decision-making agents for cyber systems (like enterprise network, industrial control systems) to mitigate highly complex cyber attacks. The development of an ACD system could be conducted through a combination of different types of operations. This includes the development of automated blue-teaming agents within Autonomous Cyber Operations Gyms as a mode of terrain (to replicate real-world cyber systems), where automated red teaming agents are used to validate, develop and strengthen the automated blue team agents for an overall goal of their real-world development. 

\subsection{Automated Cyber Defence Importance within National Strategy Documents}
\label{subsec: strategy}
Private and government-based organisations have made it clear that AI will soon be forefront within cybersecurity in terms of detecting and responding to attacks. Table \ref{tab: national-strategy-papers} elaborates the importance of ACD within different countries.


\begin{table}[b]
\resizebox{\textwidth}{!}{\begin{tabular}{c l l }
\hline
\multicolumn{1}{l}{\textbf{Country/Alliance}} & \textbf{Department/Strategy}                   & \textbf{Reference to ACD}                                                                                                                                                                                                                                                                                  \\ \hline
\multirow{3}{*}{\textbf{Australia}}             & Department of Defence \cite{australia-gov-def-2022}                       & \begin{tabular}[c]{@{}l@{}}Suggests the need to expand cybersecurity \\ skills and integrating AI into it. DoD is \\coordinating research and investment in AI \\capabilities to strengthen capability across the \\information and cyber domains.\end{tabular}                                      \\ \cline{2-3} 
                                                & AI for Decision-Making Initiative 2022 \cite{ai-for-dec-mak-init-2022}      & \begin{tabular}[c]{@{}l@{}}Aims to develop 30 more AI-based challenges \\ for researchers, including the TTCP CAGE \\Autonomous Cyber Defence Challenge to \\produce AI-based automated decision \\blue teaming algorithms for instantaneous \\response against cyber attacks.\end{tabular} \\ \cline{2-3} 
                                                & Royal Air Force of Australia \cite{raf-australia-strategy}                  & \begin{tabular}[c]{@{}l@{}}Advises continuous evaluation in which \\ decisions can be made by machines and \\ which must be made by humans.\end{tabular}                         \\ \hline
\multirow{2}{*}{\textbf{Canada}}                & National Cybersecurity Strategy \cite{canada-nat-cy-sec-strategy}             & \begin{tabular}[c]{@{}l@{}}Specifically mentioned the importance of \\ defence and security applications with \\ autonomous decision
support\end{tabular}                                                                                                                                                  \\ \cline{2-3} 
                                                & Defence Research and Development \cite{dondo-canada-strategy}              & \begin{tabular}[c]{@{}l@{}}The publication suggests that a combination \\ of deep learning and RL algorithms for accurate \\ identification of evolving threats, and then \\ recommend or execute an appropriate course \\ of action.\end{tabular}                                                         \\ \hline
\multirow{2}{*}{\textbf{United Kingdom}}        & Defence Artificial Intelligence Strategy \cite{DAIS}       & \begin{tabular}[c]{@{}l@{}}Discusses the new risks from AI-Enhanced \\ Cyber Threats which operate \\ at speeds and at scales preventing actions \\ by human operators in a timely manner.\end{tabular}                                                                                                    \\ \cline{2-3} 
                                                & Government Cybersecurity Strategy   \cite{GCSS}          & \begin{tabular}[c]{@{}l@{}}Described AI as an emerging technology \\ to focus on. Proposes to explore AI in the \\ context of detecting malicious activity \\ and in some cases to “enable automated \\ response to threats”\end{tabular}                                                                  \\ \hline
\multirow{2}{*}{\textbf{NATO}}                  & Cooperative Cyber Defence Centre of Excellence \cite{NATO1} & \begin{tabular}[c]{@{}l@{}}Suggest the need for Nation States to \\ adopt and explore AI-enabled Cyber \\ Defence.\end{tabular}                                                                                                                                                                            \\ \cline{2-3} 
                                                & NATO AI Strategy \cite{NATO2}                              & \begin{tabular}[c]{@{}l@{}}The strategy includes "collaboration on AI\\ technologies for Cyber Defence.\end{tabular}                                                                                                                                                                                       \\ \hline
\end{tabular}
}
\caption{National Strategy Paper's on ACD}
\label{tab: national-strategy-papers}
\end{table}

\newpage

\subsection{Automated Cyber Defence Requirements}
\label{subsection: requirement-analysis}

The North Atlantic Treaty Organisation (NATO) outlined requirements for Autonomous Cyber Agents by producing a reference architecture and technical roadmap, AICA \cite{aica}. A specific part of the document focuses on the strategic deployment and the ethical concerns on the battlefield of autonomous agents. The key points in AICA relevant for this paper have been included along with additional ACD requirements in a summarised requirement analysis for ACD in Table \ref{tab:req-for-acd} below. The table further elaborates essential requirements of automated red and blue agents \textbf{(A)} along with ACO Gyms requirements \textbf{(G)} which could allow the usage of automated red and blue agents. The requirements within this table should act as a checklist for researchers within ACD, allowing for the development of eventual deployment of ACD operations within real-world networked systems.

\begin{table}[]
\centering {%
\resizebox{\textwidth}{!}{\begin{tabular}{l l }
\hline
\textbf{Requirement}                                                 & \textbf{Summary}  \\ \hline
Generalisation                                                       & \begin{tabular}[c]{@{}l@{}}
\\ - \textbf{(G.1.1)} ACO Gym will need to generalise to new settings and have the ability to seamlessly \\add components
\\ - \textbf{(G.1.2)} ACO Gym would need to be able to add different types of agents.
\\ - \textbf{(G.1.3)} Networked system training-testing must promote transfer from  \\ simulation to a real world design, including aspects like matching real networked system latency \\ operations delays within networked systems. Examples include a hybrid of simulation and emulation \\ within training-testing strategies \\
- \textbf{(G.1.4)}  ACO Gym must have capability of scaling the network to larger sizes without \\ configuration issues\\ 
- \textbf{(A.1.1)} Automated agent will need to generalise their decisions relevant to the agent type it represents\\ 
- \textbf{(A.1.2)} Automated agent will have to generalise and adapt to structural changes within \\ the ACO Gyms (addition and removal of subnets and endpoints)\\ 
- \textbf{(A.1.3)} Automated red and blue agents must be designed to sustain their high performance \\ from simulation to real-world.\end{tabular} \\ \hline

\begin{tabular}[c]{@{}l@{}}  High Level\\ Decision-Making\end{tabular} & \begin{tabular}[c]{@{}l@{}}
- \textbf{(G.2.1)} ACO Gyms must be designed to explain their state after specific events occur \\ within the networked system.\\ 
- \textbf{(G.2.2)} ACO Gyms will need to be framed into MDP/POMDP format in order to allow for \\ automated decisions to be made.\\ 
- \textbf{(A.2.1)} For planning and collective response plans, sequential algorithms will need to be considered.\\ 
- \textbf{(A.2.2)} AICA reference architecture argues that both Game Theory and Artificial Intelligence \\ would be suitable for implementation within ACD.\\ 
- \textbf{(A.2.3)} The designed automated agents will require a "deep" architecture to sustain the \\ complexity of the ACO Gyms\\ 
- \textbf{(A.2.4) } Additionally, agents will need to be able to be explainable \cite{xai-robustness-hamon,brundage2020toward, kim-xai}, i.e. justify their \\ real-time decisions made in order for them to be  operational within real-world networked systems.\end{tabular}                                                                      \\ \hline
Learning                                                             & \begin{tabular}[c]{@{}l@{}}
- \textbf{(A.3.1)} AICA \cite{aica} opens up on the possibility of enabling continual learning within ACO Gyms \\ 
- \textbf{(A.3.2)} But also argues the importance of training-testing approaches\end{tabular} \\ \hline
\begin{tabular}[c]{@{}l@{}} Multi-agent\\ Collaboration\end{tabular}  & \begin{tabular}[c]{@{}l@{}}
-  \textbf{(G.4.1)} ACO Gyms must be designed in a way to allow for multi-agent reinforcement \\learning (MARL) to operate\\ 
-  \textbf{(A.4.1)} Multi-Agent System representations would be required to train the automated agents and for \\ action/strategy negotiation. \footnote{https://github.com/cage-challenge/cage-challenge-3}. AICA combined with a MARL survey produced by \cite{wong2021multiagent}, suggests utilising \\combinations of communication approaches and centralised training \& Decentralising Execution \\solutions at a bear minimum.\end{tabular}
\\ \hline

\begin{tabular}[c]{@{}l@{}}Research\\ Collaboration\end{tabular}     & 
\begin{tabular}[c]{@{}l@{}}A requirement is the need to explain and collaborate with other researchers within ACcD \cite{burke-et-al-turing} \\ that coincides with ACD. Thus:\\ 
- \textbf{(G.5.1)} ACO Gym must be open-source for researchers to contribute further to implementations\\ 
- \textbf{(G.5.2)} Documentation for ACO Gyms must be available for further development of gyms \\ and ease of research and implementation of automated agents within them\end{tabular}                                                                                          \\ \hline

Resilience                                                           & \begin{tabular}[c]{@{}l@{}}
The AICA reference architecture highlights the need for resilience against differing malware samples\\ and other algorithmic attacks. Therefore:\\ 
- \textbf{(G.6.1)} ACO Gyms must be designed to allow for automated red agent to adversarially train \\the automated blue agent to reduce the number of incorrect actions\\
- \textbf{(A.6.1)} To improve performance of automated blue team agent (the sole purpose of ACD),\\ adversarial training through an automated red agent must be encouraged.\\ 
- \textbf{(A.6.2)} Automated red agents must be provided with a wide variety of cyber attacks (specified \\within the MITRE ATT\&CK framework)\\ 
- \textbf{(A.6.3)} along with a variety of algorithmic attacks \cite{hoffman-et-al-algorithmic} to address systems vulnerabilities.\\ 
- \textbf{(A.6.4)} Automated blue and red agent must be able to launch deception defence and attacks respectively.\end{tabular}                                                                                                                                                                                            \\ \hline
\end{tabular}
}
\caption{Requirement for ACD}
\label{tab:req-for-acd}
}
\end{table}
\newpage
\section{ACD algorithms used within Custom ACO Gyms}
\label{sec: acd-custom-aco-gym}

As mentioned in the section \ref{subsection: requirement-analysis}, a typical ACD system comprises of a mode of terrain i.e. networked system, which possesses the provision to allow automated red and blue team game-playing scenarios. Recent publications within ACD have utilised automated decision-making algorithms such as Game Theory (GT), Machine Learning (ML) and Reinforcement Learning (RL) for automated blue and red teaming within custom ACO Gyms. \\
\\
ML-based solutions (along with RL-based solutions \cite{Sewak_2022_rl, nguyen2021deep,ren-et-al-rl}) have also been utilised solely for quick incident and intrusion response over the years \cite{gore-et-al-ml-ir, nila-et-al, veksler-sdl-cyber}. Specifically, Zago et al \cite{zago-et-al-ml} utilise ML techniques to analyse, detect and react against existing and upcoming cyber threats, including botnets. The proposed approach combines unsupervised and supervised approaches to create a scalable detection and reaction framework willing to decrease the error rate as well as increasing the efficiency in terms of computational resources. The approach uses dimensionality reduction algorithms and uses publicly available datasets for intrusion detection for its implementation. While sole ML-based implementations like this allow the mitigation of specific types of attacks, they do not cater to the rapid response of zero-day attacks due to their single-step response characteristic. Additionally, like zero-sum GT-based solutions, their performance does not scale to larger enterprise networks due to the algorithms not being complex enough to generalise state spaces further away from the scenario in operation. Cam et al \cite{cam-et-al-rl} also highlight how most ML-based solutions (which include supervised and unsupervised learning algorithms) provide solutions to a single-step learning problem, a feature of the algorithm that makes it infeasible for implementing it as ACD-based solutions within networked systems. Therefore, the publications selected for this section focus on sequential response that is required for automated agent to stop cyber attacks within an overall networked system. \\
\\
The rest of this section provides an overview of the recent publications within automated response for blue and red teaming respectively within custom networked systems, and analyses the publications based on their automated agents and custom ACO Gyms through the Requirement Analysis in section \ref{subsection: requirement-analysis}.

\subsection{Automated Blue Team Solutions}
The automated blue agent within a network system must be perpetually vigilant to defend the entire attacker surface in real-time, while the attacker only needs to succeed once within a single location. Due to this assymmetric scenario between cyber attackers and defenders, the defender with limited resources cannot afford to prepare for all possible attacks.\\
\\
The problem area in focus for this subsection is the mitigation of Posture-related vulnerabilities (PrV) i.e. the defender must be perpetually vigilant to defend the entire attacker surface in real-time, while the attacker only needs to succeed once within a single location within the networked system. Due to this disadvantage in security posture, the defender with limited resources cannot afford to prepare for all possible attacks. \\
\\
Table \ref{tab: automated blue teaming} below evaluates the automated blue teaming solutions along with their custom ACO Gyms that were published within literature.

\newpage
\renewcommand{\arraystretch}{1.5}

\begin{table}[]
\resizebox{\textwidth}{!}{\begin{tabular}{l l l l l l l l l l l l l l l l}
\hline
                                           & \multicolumn{15}{c}{\textbf{Automated Blue Team Custom Networked System Publications}}                                                                                                                                                          \\ \hline
\multicolumn{1}{c}{\textbf{Requirements}} & \cite{zonouz-et-al-rre} & \cite{booker-et-al} & \cite{huang-et-al-pbne} & \cite{ni-et-al-game-theory} & \cite{malialis-et-al} & \cite{gao-et-al-ddos-rl} & \cite{eghtesad-et-al} & \cite{chai-et-al-ddos} & \cite{chowdhary-et-al} & \cite{cam-et-al-rl} & \cite{wang-et-al-rl} & \cite{gao-et-al-dqn} & \cite{walter2021incorporating} & \cite{silva2022alphasoc} & \cite{roberts2020deep}\\   \hline \hline
\textbf{A.1.1} & +     &           &       & + & +       & +    & +        & +           & +           & +        & +                  & +               &     &  +  &   +       \\

                                    \hline
\textbf{A.1.2} &        &           &       & +    &       & +        & +    & +        & +    &           & +          & +          &          & +     &  +                                         \\

                                    \hline
                                    
\textbf{A.1.3} &        &       &       &       &       &       &       & +        &       & +        & +  &               &               &       &                                                                         \\ 
                                    \hline
                                    
\textbf{A.2.1} & +     & +    & +    & +    & +    & +    & +
& +    & +    & +    & +          & +      & +    & + &    +                                 \\ 
                                    \hline
                                    
\textbf{A.2.2}`& +     &           &       & +        & +        & +        & + & +        & +        & +        & +          & +      & +           & +  &    +             \\ 
                                    \hline
                                    
\textbf{A.2.3} &        &       & +        &           &           &           &           &           & +  & +       &           & +          &        &   & +                                                        \\ 
                                    \hline
\textbf{A.2.4}  & +       & +      &         &           &           &           &           &           &         &        &           &           &     &      &                                                        \\ 
                                    \hline
\textbf{A.3.1}  &        &       &         &           &           &           &           &           &         &        &           &           &        &        &                                                   \\ 
                                    \hline
\textbf{A.3.2} &    +    &  +     & +        &  +         & +         & +          &    +       & +          & +        & +       & +          &    +       & +    &    +    &  +                                                    \\ 
                                    \hline
\textbf{A.4.1}  &           &           &           &           & +           &           & +          &           & -         & -         &           &       &    &       &                                                                    \\ 
                                    \hline
                                    
\textbf{A.6.1}  &           &           &           & +            &           & +            &           &           & +            &           &           &           &          &         &                    \\ 
                                    \hline
                                    
\textbf{A.6.2}      & +        & +        &           &           &           &           &           &   &           &           &           &           &     &        &                                                  \\ 
                                    \hline
                                    
\textbf{A.6.3}      &           &           &           &           &           &           &           &           &           &           &           &           &          &          &                                                   \\ 
                                    \hline
                                    
\textbf{A.6.4}      &           &           &           &           &           &           &           &           &       &           & +          & +          &            &          &                                                     \\ 
                                    \hline
                                    \hline

\textbf{G.1.1}      &+     & +          & +          & +        &           &  +         & +          &           & +          &     & +          & +          &               &        +    &   +                                               \\ 
                                    \hline
                                    
\textbf{G.1.2}      &           &           & +          & +           &           & +          & +           &           & +      &           &           &           &         &      +      &      +                                                \\ 
                                    \hline

\textbf{G.1.3}      &           & +          &           &           & +          &           &            &           &        & +          & +          &           &            &          &                                                     \\ 
                                    \hline
                                    
\textbf{G.1.4}      & +          & +          & +          & +          &           &    +       & +           &           & +       &           &           & +          &            &     + &    +                                                     \\ 
                                    \hline
                                    
\textbf{G.2.1}      &       & +    &       &       &         &       &       &       &       &       &       &         & +           &   +       &   +                                                                          \\ 
                                    \hline
                                    
\textbf{G.2.2}      & + & +       & +          & +          & +          & +          & +          & +          & +      & +          & +          & +          & +         &                    +   &     +                                     \\ 
                                    \hline
                                    
\textbf{G.4.1}      &           &           &           &           & +         &           & +          &           & +      &           &           &           &               &          &                                                  \\ 
                                    \hline

\textbf{G.5.1}   &           &           &           &           &         &           &           &           &       &           &           &           & +    &       &                                                               \\ 
                                    \hline
                                    
\textbf{G.5.2}   &           &           &           &           &          &           &           &           &      &           &           &           & +  &          &                                                              \\ 
                                    \hline

\textbf{G.6.1}      &           &           & + & +           &           &           & +          &           & +      &           &           &           &     &        &                                                              \\ 
                                    \hline
                                    
\end{tabular}
}
\caption{Automated Blue Team Solutions within custom networked systems}
\label{tab: automated blue teaming}
\end{table}

Table \ref{tab: automated blue teaming} shows relevant ACD automated blue teaming publications within networked systems designed solely for their respective automated blue team agent implementations. The table highlights most publications meeting requirements A.1.1, A.1.2, A.2.1, A.2.2. This is specifically because most publications highlight the need for a sequential blue agent response \cite{cam-et-al-rl}, as opposed to single-shot blue agent responses that are not feasible to defend the systems against modern day cyber attacks. This is further shown by all publications framing the problem as a MDP/POMDP (G.2.2), which allows automated agents to take sequential response through the transitioning of states, that signify a combination of actions taken within specific nodes of a networked system. However, while the requirements of A.1.2 are met within the specific publications, they are simulation based networked system implementations, which means that the system does not completely represent the complexity of configuration changes of the real-world networked systems. This is specifically highlighted in A.1.3 requirement which is not met by most publications in Table \ref{tab: automated blue teaming} that only test their algorithms within simulated networked systems. Most publications did not meet A.2.3 that is required within complex networked environments for appropriate generalisation of long-term actions for the agent. Only Deep RL (DRL) implementations were able to fill this requirement, making them more suitable. Dhir et al \cite{dhir} also suggested the use of Causal Inference algorithms \cite{kaddour-causal,peng-causal, zecevic-et-al-causal, rezende-et-al-causal, gasse2021causal} that could maintain their performance within ACO Gyms. Most publications in Table \ref{tab: automated blue teaming} also do not meet explainability requirement of A.2.4, which is primal for utilisation of any automated agents within SOC environments, in which such agents will need to be certified before they are in operation. Only 2 of the selected publications met A.4.1, in which both publications implemented automated response against specific cyber attacks (i.e. DDoS, as opposed to an agent that could detect and respond to a variety of cyber attacks). Such requirement is highlighted in the form of A.6.1 and A.6.2, which suggests the need to continually develop the knowledge base of the automated blue agent through adversarial training against a variety of cyber attacks. Moreover, the lack of implementations that fill the A.4.1 requirement also hinders the development of automated blue agents against algorithmic attacks mentioned in A.4.3, an area in which no publications highlighted in Table \ref{tab: automated blue teaming} have concentrated on. \\
\\
Requirement A.6.4 in the context of automated blue teaming refers to defender agents which have the capacity to strategically launch deceptive elements that enhance the defence of a networked system through an increase in threat detection functions. Applications of Cyber Deception in literature seek to integrate high-fidelity deceptive assets into existing infrastructures with the purpose to mislead or slowdown adversaries and ultimately thwart their cognitive processes. These assets are typically encapsulated inside virtual environments that resemble their physical counterparts; and have two overall aims: first, the defence of a system through the enhancement of threat detection functions such as lures and decoys, and second, the ability to misdirect and quarantine attackers to support the gathering of Cyber Threat Intelligence (CTI). Deception-based Cyber Defence (DCD) platforms counter classic attacker-defender asymmetries by executing and maintaining preventative cybersecurity tools that, unbeknown to an adversary, obfuscate the true security posture of a network. In fact, the use of DCD is becoming an increasingly prudent choice in the mitigation of PrV(s) on the account that adversaries must ‘minesweep’ through a sea of supposed vulnerabilities in order to execute a successful cyber attack. Wang et al \cite{wang-et-al-rl} and Ghao et al \cite{gao-et-al-dqn} both consider the notion of combining the use of intelligent algorithms with dynamic deployment strategies in order to analyse adversary behaviour. Both solutions succeed in training a blue agent to select optimal deployment strategies but fall short of many generalisation and resilience-based requirements due to banding together of attackers with the associated environment. As previously mentioned, solutions such as \cite{gao-et-al-dqn} which incorporate DRL typically meet the high-level decision-making requirement A.2.3. The use of DRL in this instance is sensible because the authors are aware of the impact that general attacker-defender scenarios have on the space complexity of typical RL algorithms. This is because Deep Neural Networks (DNNs) are introduced to make policy-based deployment decisions without the need to manually engineer the state space. In the context of ACD, determining a reward path through the trial and error of all possible states can often converge to computational intractability as the scale of the network environment grows; thus, by harnessing the predictive element of a DNN, knowledge becomes generalised by approximating each Q value rather than storing and looking up every distinct state. The authors in \cite{gao-et-al-dqn} utilise online learning to update defence models with newly collected attack information, although this is of a ‘non-continual’ variety, meaning continual learning techniques have not been implemented to address concerns regarding catastrophic interference, thereby failing to meet requirement A.3.1. Leveraging the approximations of DRL, Li et al \cite{li-drl} proposes an optimal defensive deception framework by creating System Risk Graphs (SRG) which model adversary actions. The attack models are then used to train a DRL agent to generate optimal deployment strategies within micro-service architectures. Incorporating defensive deception into container-based cloud environments is sensible as, like the diversity and scale of typical OT networks, the virtualisation of technology and the dynamism of container services exposes a glut of additional attack vectors to an already overwhelming issue. Through the intelligent deployment of deceptive assets, the expanding threat surface can be maintained and prevented. The authors highlight the issue of scalability when modelling network environments and threat models as high-dimensional input spaces, implementing a DRL framework that scaled up to 60 nodes. In a different light, Walter et al \cite{walter2021incorporating} draws attention to the prospect of augmenting ACD environments with defensive cyber deception components by adapting the source code of an existing open-source ACO Gym called CyberBattleSim \cite{cyberbattlesim}. This paper falls short of many requirements as the solution does not necessarily create a dedicated blue agent. Instead, the aim of the paper was to gain insight by observing the impact of active cyber deception on attacker behaviours which can ultimately inform automated blue teaming agents. \\
\\
In terms of the requirement of networked systems within the publications mentioned in Table \ref{tab: automated blue teaming}, G.1.1 and G.1.2 were met within most simulated networked system publications. However, as mentioned previously, simulated systems do not represent the real-world systems accurately, hence the reason why very low number of mentioned implementations are able to meet the G.1.3 requirement. Similar to the requirement A.4.1, G.4.1 is an area in which networked systems will need to be developed in order to facilitate the inclusion of automated agents. Areas of research development also include G.4.1 and G.6.1, in which networked systems will need to be designed to allow such requirements. 
\subsection{Automated Red Team Solutions}

The existing literature on automated red teaming solutions can be split into three categories: assistance to security analysts with attack planning, penetration testing or red teaming “automation”, and red agent research conducted in gym environments. The later categories relate closely to ACO goals/objectives, whilst the former is an intermediary step towards it. \\
\\
The attack path planning category utilises scanning information outputted from penetration testing tools such as Nmap \cite{nmap} or Nessus \cite{nessus} to design a POMDP (G.2.2) representing a corporate network. The Common Vulnerability Scoring System (CVSS) scores \cite{cvss} from vulnerability scans are then utilised to define the transition probabilities. \cite{gangupantulu2021crown} also utilised the CVSS scores to inform the rewards (landing on the host as an administrator for instance). Researchers then utilise RL algorithms (A.2.2) on these environments to reach set objectives (while adding negative penalties at each step to avoid loops). For example, \cite{gangupantulu2021crown} and \cite{chowdhary2020autonomous} utilised this approach to generate action plans to assist a human expert in reaching testing objectives with the DQN algorithm (A.2.3). Finally, it should be noted that tools such as Bloodhound \cite{bloodhound} offer attack path planning focusing on Active Directory weaknesses, without utilising ML. \\
\\
To automate penetration testing, one can extend the RL game defined in the paragraph above to incorporate actions of penetration testing or red teaming tools (A.6.2). In fact, \cite{zhou2019nig} did so to automate penetration testing with the Metasploit framework \cite{metasploit}, whereas \cite{maeda2021automating} utilised the PowerShell Empire framework \cite{empire} to automate post exploitation activities. Furthermore, researchers have analysed specific tasks of red teaming and attempted to automate them. For example, \cite{kujanpaa2021automating} automated privilege escalation through RL. One could envision multiple cells of the MITRE ATT\&CK matrix \cite{mitre} being automated in this fashion, such as defence evasion as seen in \cite{fang2019evading}. \\
\\
Given that research into RL for automated red team solutions can be abstracted into simulated environments (described in further detail in ACO Gyms, G.1.3), the literature also comprises of such research. For example, \cite{sultana2021autonomous} build Deep RL agents in the Network Attack Simulator Gym \cite{nasim}. The authors trained agents in five different scenarios of varied sizes and complexity, which were built with the PPO and DQN algorithms. They trained them on smaller scenarios to see how they performed in the larger ones at testing time, where PPO seemed to generalise slightly better. Given the exponential growth in action sets, researchers have begun analysing the use of Hierarchical RL in this setting, in fact \cite{tran2021deep} did so in the CyBORG Gym environment \cite{standen2021cyborg} where they proposed a Hierarchical DQN algorithm. Research in the open-source gyms are summarised in \ref{tab:automated-red-team-in-gyms}.\\
\\
Finally, it should be noted that GT Models (A.2.2) have also been explored (an example is provided by \cite{colbert2020game}), but in this case they are utilised to aid decision makers, such as in cyber war gaming.

\renewcommand{\arraystretch}{1.5}

\begin{table}[]
\begin{tabular}{l l l l}
                                           & \multicolumn{3}{c}{\textbf{Automated Red Team Custom Networked System Publications}}                                                                                     \\ \hline \hline
\multicolumn{1}{c}{\textbf{Requirement}} &  \cite{maeda2021automating}& \cite{kujanpaa2021automating} & \cite{gangupantulu2021crown}  \\ \hline \hline
\textbf{A.1.1}                             &               &                   &                \\ \hline
\textbf{A.1.2}                              &               &                   &                \\ \hline
\textbf{A.1.3}                             &               &          +         &                \\ \hline
\textbf{A.2.1}                              &    +           &        +           &    +            \\ \hline
\textbf{A.2.2}                             &     +          &         +          &      +          \\ \hline
\textbf{A.2.3}                              &    +           &       +            &      +          \\ \hline
\textbf{A.2.4}                             &               &                   &                \\ \hline
\textbf{A.3.1}                             &               &                   &                \\ \hline
\textbf{A.3.2}                              &               &                   &                \\ \hline
\textbf{A.4.1}                             &               &                   &                \\ \hline
\textbf{A.6.1}                            &               &                   &                \\ \hline
\textbf{A.6.2}                            &               &                   &                \\ \hline
\textbf{A.6.3}                             &               &                   &                \\ \hline
\textbf{A.6.4}                             &               &                   &                \\ \hline \hline
\textbf{G.1.1}                              &    +           &                   &       +         \\ \hline
\textbf{G.1.2}                             &               &                   &                \\ \hline
\textbf{G.1.3}                              &       +        &      +             &     +           \\ \hline
\textbf{G.1.4}                             &        +       &                   &      +          \\ \hline
\textbf{G.2.1}                             &              &                   &                \\ \hline
\textbf{G.2.2}                             &        +       &     +              &      +          \\ \hline
\textbf{G.4.1}                             &               &                   &                \\ \hline
\textbf{G.5.1}                             &               &                   &                \\ \hline
\textbf{G.5.2}                             &               &                   &                \\ \hline
\textbf{G.6.1}                             &               &                   &                \\ \hline
\end{tabular}
\caption{Automated Red Team solutions within custom networked systems}
\label{tab:automated-red-team-custom}
\end{table}
\newpage
\section{Autonomous Cyber Operations Gym}
\label{sec: aco-gyms}

As shown in the previous section, the lack of common open-source ACO Gyms prevent the possibility for a separate accelerated development of automated blue and red agents (and ACO Gyms). This section provides a detailed overview of literature that have recently developed ACO Gyms along with the automated agents developed and published within literature and websites. Such ACO Gyms are simulated and/or emulated networked systems designed specifically for the development of automated blue and red team solutions. Given the availability of several resources, different publications have produced different strategies for training and testing environments, algorithm development type, and the types of cyber attacks possible.  

\subsection{Training strategies}
The most common approach to train and test an agent involves validating its policies on the same environment in which it was trained. This applies to both simulated and emulated environments. Unfortunately, this strategy prevents making statements on the automated agent's ability to generalise (i.e. meet requirement A.1.1, A.1.2 and in Table \ref{tab:req-for-acd}). Additionally, the automated agent will not be able to fully utilise the benefits of using different types of environments (i.e. simulation for scalability and emulation to delve closer to realism) and meet requirement G.1.3. \\
\\
Several research papers have strode to make progress in the domain of generalisation. For example, \cite{sultana2021autonomous} built Deep RL agents in the Network Attack Simulator Gym \cite{nasim}; a simulated environment to conduct research in automated penetration testing. Automated agents were trained in five different scenarios (encompassing subnets, hosts, vulnerabilities) of varied sizes and complexity, where the authors adopted both the PPO and DQN algorithms. After training the automated agents on scenarios of lower complexity, the impact on performance in larger complexity scenarios was experimented with, where the PPO provided superior generalisation. The cutting-edge platforms built to conduct research in ACD designed by \cite{farland-MolinaMarkham-et-al}, \cite{standen2021cyborg} or \cite{li2021cygil} all involve a simulated environment to train agents in a time efficient manner. In addition, emulations of the environment can be spun up on cloud providers with services running, actual malware performing malicious actions and automated blue agents with abilities to close ports or remove infections (mapping to the action spaces of the simulation). Another approach involves “real world” testing after training is performed in a simulated environment. One example worth mentioning are task specific agents, for example, \cite{kujanpaa2021automating} enumerated all possible privilege escalation techniques from the MITRE ATT\&CK matrix \cite{mitre} and built an agent with DQN to perform this task. In order to speed up the learning process, they trained their agent in simulated environment built with Python and then conducted their testing in the “real world” (a Windows Virtual Machine).
They measured its performance based on how many steps were needed to escalate privileges, for some cases/vulnerabilities, the automated agent outperformed human experts.

\subsection{Existing Autonomous Cyber Operations Gyms}
For the acceleration of research within the domain of automated red and blue teaming agents within networked systems, open-source networked systems, or Autonomous Cyber Operations Gym (ACO Gyms) will be required. The provision of ACO Gyms will allow researchers to streamline their focus on meeting the automated agent based requirements in Table \ref{tab:req-for-acd}. In addition, this allows researchers to also focus on developing more open-source ACO Gyms that meet the networked system requirements in Table \ref{tab:req-for-acd}. Below is a review of existing environments which are designed for cybersecurity research. The review begins with providing an overview of the existing environments that are simulations, and then delves into other closed-source emulated (and other simulated) environments that have been published. Each part compares ACO Gyms amongst the other open-source/closed-source ACO Gyms using the requirement analysis for ACO Gyms.

\subsubsection{Open-source Gyms}

Firstly, The Cyber Battle Sim \cite{cyberbattlesim} (CBS) environment is created for training automated red agents that focus on the lateral movement phase of a cyber-attack in an environment that simulates a fixed network with configured vulnerabilities. The red agent utilises exploits (specific code that remotely access a network and gain elevated privileges, or move deeper into the network) for lateral movement while a pre-defined blue agent aims to detect the red agent and obstruct access. The CBS environment can define the network layout and the list of vulnerabilities with their associated nodes. In CBS, the modelled cyber assets capture OS versions with a focus to illustrate how the latest operating systems and up-to-date patches can deliver improved protections. The implementation can also be extended due to its design for “blue agent” training. In fact, \cite{walter2021incorporating} have done so to incorporate blue team deception into the environment. The developers ensured sufficient complexity exists in the environment to abstract the cells of the MITRE ATT\&CK matrix \cite{mitre} for vulnerabilities (to be exploited by red agents to get rewards). Overall, the documentation is sufficient to create new scenarios/networks, tweaking reward functions (values of compromised services and costs of exploitation) and adding vulnerabilities to services. While this allows users to extensively experiment with the environment, the code only exists for implementation within a simulated domain, thereby questioning the realism of the environment. \\
\\
The Gym IDS Game \cite{hammar2020finding} is a simplistic Markov game built upon the OpenAI gym environment. The attacker has two types of available actions.
\begin{itemize}
    \item A reconnaissance action
    \item Or an attack of type 1...m 
\end{itemize}
The defender also has two types of actions at his disposal.
\begin{itemize}
    \item A monitoring action
    \item Or a defensive action of type 1...m 
\end{itemize}
Different scenarios exist for either training a blue or red agent (or both). Unfortunately, the gym environment is overly simplistic and only provides a simulated environment, meaning that, like CBS, it also provides low realism. Similarly, to the Gym IDS Game described above, the Gym Threat Defence gym \cite{miehling2015optimal} is also a simulation-based system with a POMDP set-up. However, in this case, the authors have designed it as a purely defensive game where the defender has four different available actions.
\begin{itemize}
    \item No action
    \item Blocking a service
    \item Disconnecting a machine
    \item Performing action 2 and 3 in parallel
\end{itemize}
One can define the probabilities of detection for each node, the attack probabilities, the spread probabilities, and the initial state.\\
\\
The Network Attack Simulator environment \cite{nasim}, is purely built for training red agents (as there is no blue agent) to test AI systems in penetration testing tasks. This environment is built upon OpenAI gym and allows the ability to create scenarios by defining the number of hosts, services, the observability mode (fully observed for instance) and the asset criticality of the hosts in question. Finally, one can decide the vulnerabilities present on the network and define the cost of actions (cost of a subnet scan for instance). The red agent can select from seven different action types: Exploitation, Privilege escalation, Service scan, Operating system scan, Subnet scan, Process scan and No action. The goal of the project is to train red agents in performing penetration tests against simulated scenarios, while no blue agent interferes with the environment. While this implementation provides the ability to bolster research progress within AI-based red agent training, it to only provides simulations like CBS, Gym IDS Game and Gym Threat Defence, reducing the realism of the implementation. \\
\\
Similar to the environments mentioned, the Optimal Intrusion Response Gym \cite{hammar2021learning} is a Markov game built upon the OpenAI Gym libraries. The environment comprises of a simulated enterprise network with 6 subnets, with several hosts, each comprising of an IDS. 
Unfortunately, the game is overly simplistic for our use case as the defender can only select from two actions.
\begin{itemize}
    \item ”Stop” will block the gateway. This will degrade the IT service and has a cost associated with it. However, it will also ensure the infection is contained.
    \item “Continue” is a non-action.
\end{itemize}
After doing some simulations/tests, \cite{hammar2021learning} discovered that the blue agents they trained are more likely to ``Stop" earlier when facing a stealthy attacker than against a noisier one. \\ 
\\
The CyBORG environment \cite{standen2021cyborg} is designed specifically for training blue agents. However similarly to CBS, it can simply be extended for red teaming use cases. The environment allows training and testing in simulated and emulated environments respectively. The simulated environment comprises of an agent interacting with a scenario modelled in a finite state machine (FSM), in which each state represents a systems and networks. An action satisfying a respective pre-condition is required to move from one state to another. The state also provides specific details such as the creation and deletion of individual files, or the making or breaking of network connections. All combined, an ideal training environment is generated for both the defender and adversarial agent. Once the automated agent is trained, it is ready tested in the emulator, which comprises of AWS virtual machines to create a high fidelity cybersecurity environment in which the automated agent interacts with. The purpose of the environment is to act as a platform for research in ACD, whereby challenges are open to the public. Namely, the TTCP Cage Challenge 1, 2 and 3. The challenges are enterprise network environments with ascending complexity (in terms of the observation and action space for the red and blue agent).\\
\\
In the TTCP CAGE challenge 2 (the most recent challenge), the action sets for the blue agent are exhaustive.
\begin{itemize}
    \item Remove - removes malware from a host.
    \item Restore - if malware has elevated privileges it cannot be removed, and the host must be restored from backup (with a cost associated with it).
    \item Analyse - monitoring does not always detect infection (5/100 times) but performing an analysis on the host will always detect it.
    \item Decoy service - sets up a decoy service on a specific host to delay and detect red agent activity (there are 7 different services available).
    \item No action - Monitoring occurs regardless of other actions.
\end{itemize}
Scenarios can be defined in YAML files (i.e network topology and asset criticality). In addition, the project comes with varying red agents utilising different strategies. Finally, the documentation is exhaustive. This environment appears to fit all our needs for experimentation and details the high-level desired actions of an autonomous blue agent.
On top of this simulated environment CybORG extends to an emulation (which is closed source), which can be spun up on AWS to validate the trained agents. 
\\
\\
YAWNING TITAN \cite{collyer2022acd} is a highly abstracted graph-based gym for training blue agents. The action spaces for both the blue and red agents do not map to realistic ones expected for cyber defence. Instead, it appears that the gym has been created to efficiently test and validate approaches/algorithms. The graph-based design also suggests it’s true purpose is to explore computationally expensive approaches involving generalisation A.1.2 as networks can be defined as functions where the YAML file determines the behaviours and spaces. Table \ref{tab:open-source-gyms} has been used to summarise all open-source ACO gyms that can be experimented with. 
\begin{table}[]
\begin{tabular}{l l l l l l l l}
\hline
                                           & \multicolumn{7}{c}{\textbf{Automated Cyber Operations Gym (Open-source)}} \\ \hline
\multicolumn{1}{c}{\textbf{Requirement}} &  \href{https://github.com/microsoft/CyberBattleSim}{CBS} \cite{cyberbattlesim}& \href{https://github.com/Limmen/gym-idsgame}{GIG}& \href{https://github.com/hampusramstrom/gym-threat-defense}{GTD} &
\href{https://github.com/Limmen/gym-optimal-intrusion-response}{OIR}&
\href{https://github.com/cage-challenge/cage-challenge-2}{CybORG} \cite{standen2021cyborg}& \href{https://github.com/Jjschwartz/NetworkAttackSimulator}{NaSim} &       
 \href{https://github.com/dstl/YAWNING-TITAN}{YT} \cite{yawning}   \\ \hline
\textbf{G.1.1}                             &       &       &       &       &     &       &    +       \\ \hline
\textbf{G.1.2}                             &      &       &       & +       &  +     &      &           \\ \hline
\textbf{G.1.3}                             &       &       &       &       & +      &       &            \\ \hline
\textbf{G.1.4}                             &      &      &     &     &       &      & +           \\ \hline
\textbf{G.2.1}                             & +      &       &       &       & +      &  +     &            \\ \hline
\textbf{G.2.2}                             & +      & +      &       & +       &  +     & +      &   +    \\ \hline
\textbf{G.4.1}                             &       &       &       &       & +      &       &        \\ \hline
\textbf{G.5.1}                             & +      & +      &   +    & +       &  +     &  +     &   +   \\ \hline
\textbf{G.5.2}                             & +      & + &       & +       &  +     & +      &     +   \\ \hline
\textbf{G.6.1}                             &       &       &       &       &      &       &          \\ \hline
\end{tabular}
\caption{ACO Gyms (Open-source)}
\label{tab:open-source-gyms}
\end{table}

\subsubsection{Closed-source Gyms}

The rest of the ACO Gyms have been analysed in Table \ref{tab:closed-source-gyms} through the requirement analysis shown in Table \ref{tab:req-for-acd}. While the ACO Gyms highlighted are not open-source, they can provide important insights within the ACD community, particularly for researchers who can take inspiration when designing or making modifications to the existing ACO gyms. 

\begin{table}[]
\begin{tabular}{l l l l l l l l l l }
\hline
                                           & \multicolumn{9}{c}{\textbf{Automated Cyber Operations Gym (Closed-source)}}                                                                                                                        \\ \hline
\multicolumn{1}{c}{\textbf{Requirement}} & \cite{insight-futoransky-et-al} & \cite{deterLab}  & \cite{smallworld-furfaro-et-al} & \cite{cyams-brown-et-al} & \cite{candles-rush-et-al} & \cite{galaxy-schoonover-et-al} & \cite{li2021cygil} & \cite{farland-MolinaMarkham-et-al} & \cite{vine-eskridge-et-al}\\ \hline
\textbf{G.1.1}      & +&          &  +& +& +&           &           & +  &                     \\ \hline
\textbf{G.1.2}      &           &          &  +&           & +& +& +& +  & +                   \\ \hline
\textbf{G.1.3}      &           &+&            & +& +& +& +& +  & +                            \\ \hline
\textbf{G.1.4}       & + & & & + & + & + &  &  & + \\ \hline
\textbf{G.2.1}      & +&+&            & +&           & +& +& +  &                             \\ \hline
\textbf{G.2.2}      &  &    & & & &+& +& +  &                                               \\ \hline
\textbf{G.4.1}      &  &    &   &  &    &  &     & &                             \\ \hline
\textbf{G.5.1}      &   &     &    &    &  &    & &     &                             \\ \hline
\textbf{G.5.2}      &   &  &   &   &    & &   &   &                             \\ \hline
\textbf{G.6.1}      &   &  &   &    & +&+ &   & + &                             \\ \hline
\end{tabular}
\caption{ACO Gyms (Closed-source)}
\label{tab:closed-source-gyms}
\end{table}
\newpage
\subsection{Combined Analysis of all ACO Gyms}

As shown in Table \ref{tab:open-source-gyms}, most authors have recognised the requirement of the seamless addition and removal of nodes and components (G.1.1). Authors also meet the requirement of the adding automated agents (G.1.2) that are able to generalise their decisions along with understanding the structural changes within the ACO Gyms (A.1.1 and A.1.2 respectively). Moreover, all publications have also understood the requirement of AI-based sequential decision-making automated red and blue agents (A.2.1 and A.2.2 respectively), and have structured the ACO Gym as a MDP in order to facilitate such agents. However, while such ACO Gyms are highly scalable (G.1.4) and allow the development of relevant automated agents, the environments utilised in all implementations are simulations to real networked systems, highlighting the lack of open-source emulated/real-world ACO Gyms (G.1.3). This results in the lack of "real-world" experience of automated agents, which will essential for utilisation within current networked systems.\\
\\
While the rest of the analysis apply to those of automated agents, the design of the current state of the ACO Gyms could be used to assess the quality of automated agents that could be designed within the ACO Gyms. Overall, only one ACO Gym (CybORG \cite{standen2021cyborg} Cage Challenge 3 \cite{cage_challenge_3_announcement}) has recognised the need for automated multi-agent algorithms (A.4.1) as automated blue team solutions.  \cite{malialis-et-al} and \cite{eghtesad-et-al} publications (specifically focusing on using RL for defending against DDoS attacks) environments could be a potential inspiration for structuring the ACO Gyms to facilitate multi-agent automated red and blue teaming collaboration (requirement G.4.1). Very few ACO Gyms facilitate adversarial training (G.6.1 and A.6.1), which could potentially utilised to strengthen the automated blue agent against a variety of cyber attacks (A.6.2). No open-source ACO Gyms currently available have recognised the need of incorporating algorithmic cyber attacks (A.6.3) within the action space of automated red agents against automated blue agents. Inspiration can be taken from a closed-source ACO Gym \cite{farland-MolinaMarkham-et-al} to incorporate algorithmic attacks such as evasion and poising of automated agents such as DRL algorithms. 

\section{ACD Algorithms within open-source ACO Gyms}
\label{sec: acd-in-aco-gyms}
Out of the open-source ACO Gyms mentioned in the previous section, several automated decision-making algorithm's have been utilised for training and testing as automated agents. The ACO Gym creators and automated blue and red team agent developers have recognised the need for DRL-based solutions within the domain due to their nature of sequential response. While many of the requirements are met through the use of DRL-based solutions, this section suggests several gaps that still exist within the design of the automated agents through currently published implementations. Such gaps will require being met before the algorithms can be deployed into real-world networked systems for cybersecurity. Out of the current ACO Gyms, only two open-source ACO Gyms have been utilised in the publications of automated red and blue agents. In addition, many algorithms have been developed and are released open-source to promote research and development within the domain. CybORG \cite{standen2021cyborg} released three challenges with simulated networked systems with varying ACO Gym complexity in terms of the actions and observation spaces. The challenges focus on the development of automated blue agents, while the development of automated red agents (comprising of two different types of cyber attacks) is also possible. NaSim \cite{nasim} authors made their code open-source for the development of automated red agents and a few publications and implementations have utilised the simulated networks for the development of such agents. 

\subsection{Automated Blue Team Solutions}
Out of the two ACO gyms mentioned above, CybORG has published its results for the challenges \cite{cage_cyborg_2022} released, and has listed and ranked the RL-based algorithms that were used in Cage Challenge 1 \cite{cage_challenge_1_announcement} and Cage Challenge 2 \cite{cage_challenge_2_announcement} (Cage Challenge 3 results will be soon released \cite{cage_challenge_3_announcement}) through performance metrics set by the authors. Several approaches taken by different teams, and multiple unique strategies that were implemented by the automated agents. The best performing approaches across the challenges have been selected in this article and have been compared against the requirement analysis in Table \ref{tab:req-for-acd}. \\
\\
From Cage Challenge 1, Team Mindrake \cite{foley-et-al-acd-cage} won the challenge and produced a Hierarchical RL algorithm that included proximal policy optimisation \cite{ppo} with curiosity. The hierarchical \cite{hierarchical-rl-hengst} component of the algorithm is utilised through a controller to take relevant action according to the type of adversary that is deployed against the automated agent (B\_line and Meander APT agent). Models are pre-trained against both adversaries separately from the training phase and are then tested by the same adversaries at random episodes. The curiosity component allows exploration within the environment in the training phase via intrinsic reward \cite{Pathak-et-al-curiousity}, improving the reward achieved by nearly double. While the automated agent was victorious within the challenge, it does not meet the requirements A.1.3, A.2.4, A.3.1, A.4.1, A.6.3 and A.6.4. This is primarily due to the availability of the actions that could be taken amidst the two adversaries, along with the variety of attacks that could be conducted by the adversaries. Additionally, the environment \cite{cage_challenge_1_announcement} cannot facilitate A.4.1. Similarly, the other three submissions also met the same requirements as the winners of the challenge. \\
\\
From Cage Challenge 2, the team from Cardiff University (with GitHub code \footnote{https://github.com/john-cardiff/-cyborg-cage-2}) won the challenge and also produced a Hierarchical PPO similar to Team Mindrake in Cage Challenge 1. However, the team utilised the availability of deception within the 2nd challenge through the selection of decoys (when required within the scenario) in a greedy manner. Using the requirement analysis, the automated agent was not able to meet the requirements A.1.3, A.2.4, A.3.1, A.4.1 and A.6.3, but met the requirement of using deception due to its availability within Cage Challenge 2. \\
\\
Overall, as shown in both challenges, variations of hierarchical PPO agents have shown most optimal performance (also suggested and algorithmically proven in \cite{wolk2022beyond}) as compared to other approaches. While the automated agents are able to generalise the moves of the two adversaries, the environment in which they were trained on did not comprise of many different types of cyber and algorithmic attacks (A.6.2, A.6.3) for the automated agents to generalise a greater pool of algorithmic attacks. To meet these requirements within this ACO Gym, future implementations could modify the ACO Gym to increase their cyber and algorithmic attack capabilities to assess the quality of generalisation of the automated agents against a greater pool of attacks. In contrast, no automated agent implementations in both challenges provided any form of explainability (A.2.4) regarding their incoming actions that they will take.

\subsection{Automated Red Team Solutions}
Unfortunately, unlike for the Automated Blue Team Solutions, no public challenges have been proposed. As a result, research has been conducted in different gyms and under varying configurations. Therefore public comparable benchmarks are lacking.
\\ \\ 
Automated Red Teaming Solutions have so far largely been performed through Reinforcement Learning in ACD gym environments such as CyBORG \cite{standen2021cyborg}, Network Attack Simulator \cite{nasim} and CyberBattleSim \cite{cyberbattlesim}, or in emulators or custom representations of IT networks. This intuitively makes sense as the problem is perfectly modelled for a Reinforcement Learning game (exploring a POMDP).  Similarly to Automated Blue Teaming solutions, the Proximal Policy Optimisation algorithm has shown to be the most successful approach. 
\\  \\
One example worth noting, involves research conducted in the CyBORG gym by \cite{standen2021cyborg} which presents the only known example of transferring a simulated red agents into an emulation. Researcher implemented  DQN  agents  in  the  CyBORG simulator. They then validated the automated agents in the CyBORG emulator (G.1.3). Most   of   the   automated agents  successfully  transferred    to    the    emulator. Those  which  didn't  likely failed   due   to   over fitting to  the  observation  in  the simulator (moving from a discrete to continuous timed observations).\\
Another example from the Nasim gym,  presents the first example of scaling generalisation (G.1.1) was conducted by \cite{sultana2021autonomous}. They
implemented  Deep  RL  agents  trained in small scenarios and validated on larger ones at testing time. Their research suggested that the Proximal Policy Optimisation algorithm  seemed  to  generalise slightly better than other algorithms.
\\ 
\\
However, it remains an open-question if such algorithms are the most appropriate, indeed there appears to be a lack of research on casual approaches in Automated Red Teaming Solutions, even though these have recently been shown to be promising for the Blue Teaming side \cite{andrew2022developing}.

\renewcommand{\arraystretch}{1.5}

\begin{table}[h]
\begin{tabular}{l l l l l l l l l}
\hline
                                           & \multicolumn{4}{c}{\textbf{Automated Red Team}}                                                                                     \\ \hline \hline
\multicolumn{1}{c}{\textbf{Papers}} & \cite{tran2021deep} & \cite{standen2021cyborg} & \cite{nguyen2020multiple} & \cite{sultana2021autonomous}   \\ \hline \hline
\textbf{A.1.1}                                           &                                        &                        &                       &                 \\ \hline
\textbf{A.1.2}                                          &                                         &                        &                       &                 \\ \hline
\textbf{A.1.3}                                         &                                        &        +                &                       &                 \\ \hline
\textbf{A.2.1}                                         &          +                             &          +              &         +              &         +        \\ \hline
\textbf{A.2.2}                                         &           +                           &         +               &        +               &          +       \\ \hline
\textbf{A.2.3}                                         &              +                        &       +                 &      +                 &         +        \\ \hline
\textbf{A.2.4}                                         &                                       &                        &                       &                 \\ \hline
\textbf{A.3.1}                                         &                                       &                        &                       &                 \\ \hline
\textbf{A.3.2}                                         &                                       &                        &                       &                 \\ \hline
\textbf{A.4.1}                                         &                                      &                        &                       &                 \\ \hline
\textbf{A.6.1}                                         &                                       &                        &                       &                 \\ \hline
\textbf{A.6.2}                                         &                                        &                        &                       &                 \\ \hline
\textbf{A.6.3}                                         &                                      &                        &                       &                 \\ \hline
\textbf{A.6.4}                                         &                                       &                        &                       &                 \\ \hline 
\textbf{Gym}                                         &                 CyBORG                      &            CyBORG            &           Nasim            &   Nasim              \\ \hline 
\end{tabular}
\caption{Automated Red Team solutions within open-source Gyms}
\label{tab:automated-red-team-in-gyms}
\end{table}
\newpage
\section{Discussion}
\label{sec: challenges}

This main purpose of this paper was identify an imminent research area, ACD, within cybersecurity in order to mitigate cyber attacks in the future. Automated response to cyber attacks will need to be addressed through the research and development of automated red and blue teaming agents that are sequential in the nature of their decision making. The development of such algorithms could be accelerated through a parallel research and development within the area of ACO Gyms. While recent advancements have developed the research area in particular directions, more challenges have been identified through the requirement analysis (Table \ref{tab:req-for-acd}) in this paper for the future development within the mentioned areas. Over 40 publications were analysed and compared through the requirement analysis in Table \ref{tab:req-for-acd}. While development of ACO Gyms and automated red and blue agent comprise of separate research and development strategies, the progress of one area is heavily dependent on the other, justifying the reasoning of having common research challenges. Since more challenges may exist within the specific requirement addressed, it is encouraged for researchers to build on this document to further address and develop areas within ACD that could further catalyse it's development into industrial use.
\subsection{Challenges and their Importance}
The direct mapping of the requirement analysis in Table \ref{tab:req-for-acd} to the publications identified as ACD has addressed that there are evident challenges that need to be filled for ACD systems to be further developed before they are implemented into real-world systems. This section outlines the areas of further research and development that were identified, and links the areas back to the specific requirements within the requirement analysis. Requirements have been added for each challenge in a descending order of importance. 

\subsubsection{AI-based Attack Robustification of Automated Blue Agents (A.6.3, G.6.1, A.6.1)}
This focuses on the area includes the robustification of Deep RL algorithms against poisoning and evasion attacks that aim to attack the algorithmic functionalities of the automated agent. While very low number of publications have focused on such attacks for Deep RL algorithms, it is evident that the future cyber attackers will implement such attacks in the future through Deep RL and neural network based research within other domains \cite{poisoning-evasion-shafahi, poisoning-evasion-zhu, rl-evasion-apruzzese, rl-attack-chen}. If this challenge is not addressed, future networked systems could be vulnerable to algorithmic attacks that could potentially take control of the automated blue agent, and eventually the entire network.

\subsubsection{Continual evolution of action space for the Automated Red Agents (A.3.1, G.6.1, A.6.1, A.6.2, A.6.3)} Red agents action spaces are constantly evolving. Indeed, new services are often added which may have vulnerabilities tied to them. In addition, “every year new exploits are found for software and so in order to be useful automated penetration testing agents will need to be able to handle a large growing database of exploits.” \cite{automatedpt-Schwartz-et-al} Therefore the red agents and the gyms they are trained in would need to consider this challenge (G.1.1). While this challenge is reliant on challenges \textit{7.1.1 and 7.1.5}, the development and addition of cyber attacks automated red agents based within a continual learning setting are yet to be explored. Failure to implement on these challenges will keep the automated blue agent agent outdated from latest cyber and algorithmic attacks.   

\subsubsection{Explainable RL (A.2.4)} 
Explainable RL is more complicated than XAI, in fact “explainability for an RL agent, while clearly a subset of XAI and with similarities to IML (Interpretable ML), has distinct characteristics that requires its explicit separation from current XAI and IML research” \cite{dazeley2021explainable}.  Indeed, the first difficulty for XRL is due to the long-time horizons which determine the decisions/actions to take. The second one relates to the models not being built off labelled training data (which would simplify explainability). Further inspiration could be taken from relevant survey papers and implementations \cite{hierarchical-rl-mitchener, hierarchical-rl-shu, xrl-ammanabrolu, xrl-lyu, xrl-peng, xrl-survey-glanois, xrl-survey-milani, xml-knowledge-graphs, madumal-et-al-xrl, puiutta-et-al-xrl, olson2021counterfactual}. Failure to address this challenge will lead to the automated blue agent not being certified by industrial employees within networked systems since the trust towards the agent will be low.

\subsubsection{Multi-agent RL (G.4.1)} 
Another research area within automated blue teaming for ACD is the utilisation of Multi-agent RL algorithms as opposed to using single RL algorithms for implementation. This will be particularly more beneficial within enterprise networks environments which are highly complex. While \cite{standen2021cyborg} authors have proposed the implementation of multi-agent RL within their third Cage Challenge \footnote{https://github.com/cage-challenge/cage-challenge-3}, more research areas could emerge with more research within this domain. Using single automated blue teaming agents will be useful, however, mistakes made by the agent within non-work hours will not be addressed unless there is another agent that evaluates the first agent and alerts it if a wrong decision is made.

\subsubsection{Cybersecurity Attack Robustification of Automated Blue Agents (A.6.2, A.6.1)}
An area for improvement for future automated blue agents (and ACO Gyms) is the implementation of more types of cyber-attacks that could occur within an enterprise network. A useful framework for this could be the use of different cyber-attacks that have been listed within the MITRE ATT\&CK matrix. Similar to software updates and patches, systems could be designed in such a way so that more attacks could be added to a knowledge base once they are listed within frameworks like MITRE ATT\&CK matrix \footnote{https://attack.mitre.org/matrices/enterprise/}. Failure to address this issue within training will lead to the automated blue teaming agent ignoring specific cyber attacks, eventually leading to a breach within the network.

\subsubsection{Robustification of Deception Techniques in Automated Blue Agents (A.6.4)}
It’s also important to highlight the necessity for research areas which utilise deception technology for ACD purposes. Their inclusion within ACO Gyms will allow the introduction more complex and proactive defensive deception techniques in order to study their effects in misdirecting and disrupting adversaries along the cyber kill chain. Existing literature rarely considers the complexity of this challenge, underlining the infancy of deception as a tool for ACD. Research that falls into this category \cite{gao-et-al-dqn, wang-et-al-rl, walter2021incorporating} typically prioritise the use of honey-x methods \cite{pawlicktaxonomy} or ‘lures’ to analyse adversary behaviours through intelligent deployment strategies. A useful framework to encourage diversity within deceptive assets is the MITRE ENGAGE matrix, which identifies numerous deception techniques that can be leveraged at different areas of ACD to optimise adversary engagement \footnote{https://engage.mitre.org/}. Failure to address this challenge deflects from the key purpose of deception as adversaries can weaponise on the homogeneity of decoys and thus magnify the asymmetry that’s ever-present between blue and red agents \ref{sec: acd-custom-aco-gym}. 

\subsubsection{Realism of ACO Gyms (G.1.3, A.3.1, A.3.2, G.1.4, G.1.1, G.1.2)} 
Another challenge within the ACO gyms is the lack of realism of most of the environments that currently exist. A metric to classify the quality of the training-testing (or continual learning) strategy as a research area is particularly important. Additionally, researchers generally would require building simulated environments and then transfer the learned policies to the real world (Sim-to-Real Transfer), this is often done in the case of robotics as pointed out by \cite{zhao2020sim}. Environments such as CyBORG \cite{standen2021cyborg} attempt to address this challenge by supporting both simulation and emulation, however, both implementations comprise of areas which do not represent real networked systems (i.e. latency delays in simulation and network scalability in emulation). In addition, IT and OT networks, unlike traditional RL tasks, are continual and ever-changing environments, which unlike most RL tasks, networks and hosts in a corporate environment are non-stationary, whereas video games in which RL have been used would not expect an agent to perform well on an entirely new map \cite{huang-et-al-reward-poisoning, berner2019dota, vinyals2019grandmaster}. Such issues must be addressed, else the agent will not recognise the mode of terrain it is operated within, leading to incorrect actions being taken by the agent.

\subsubsection{Realism of Deception Techniques (A.6.4)}
Deception fidelity is often overlooked and introduced as a part of a constraint or assumption in current literature. As virtualisation of physical assets becomes more commonplace in  context of network emulation, the implementation of Deception-based Cyber Defence (DCD) platforms must have the capability to model and simulate physical processes to maintain system fidelity and not alert attackers of its use. However, it is difficult to strike a balance between system fidelity and a sizable attack surface, particularly when considering the complexity and scale of some networked systems such as Operational Technology (OT) environments, where researchers must find methods to emulate devices in convincing ways without replicating the network in its entirety. There needs to be new methods for creating decoy profiles for assets which embody the attributes of the network component. Researchers can also look to deceptive techniques which already consider or enhance the fidelity of integrated-lures. ‘Honeyshills’ \cite{hofer2019-honeyshills} are an example as they use real components or systems and configure them to communicate with decoys to further give the impression of realism. These encourage suggestions for scaling deception methods within simulation-based networks and ultimately the move towards the emulated domain. Failure to address these problems may result in the exposure of deception to the attacker, nullifying the precedence of deception over an attacker’s inadvertence to its use. Such a contradiction cancels-out the symmetric advantage that’s provided by correctly implementing deception technology.

\subsubsection{Impact of Incorrect Action (G.6.1, G.1.3) \cite{hoffman-et-al-inc-act}} 
The above issue also leads on a gap within the ACD literature for automated decision-making agents. Appropriate evaluation and metrics will need to be explored. Additionally, approaches such as Hierarchical RL, Neurosymbolic RL and other explainable implementations \cite{hierarchical-rl-mitchener, hierarchical-rl-shu, metrics-xai} must be explored for superior forensic evaluation of the automated agents. Not addressing this area will result in the automated blue team RL agent potentially eliminating important processes within the network, which could even lead to high monetary losses.

\subsubsection{Action and Observation Spaces (G.2.1 ,G.2.2, A.2.3)} 
The first difficulty relates to the huge action and observation spaces: Existing research in ACD significantly reduces the action and observation spaces by abstracting the action spaces to a point where they may no longer be usable in the “real world”. Indeed, in a Cyber  Security  setting  where  agents  may  be  deployed  on  thousands of  hosts  (in  a  single  corporate  network),  each  with  huge  action  sets  (kill  any  process,  re-move/quarantine  any  file,  change  any  firewall  setting  etc.)   and essentially a continuous observation space, it would be challenging to sufficiently explore the space in training. This challenge applies to automated red agents also as “applying conventional DRL to automate penetration testing would be difficult and unstable as the action space can explode to thousands even for relatively small scenarios”  where “each action in PT can have very different effects such as attacking hosts in different subnets or different method of exploits”  \cite{tran2021deep}.

\section{Conclusion}
\label{sec: conc}

This article provided awareness to the area of Automated Cyber Defence by defining its terminology through research publications, government strategy reports and cybersecurity training organisations. Subsequently, the terminology allowed the segmentation of different sub-topics of research and development that exist within the area, namely, Automated Agents and Autonomous Cyber Operations (ACO) Gyms. The recognition of the sub-topics allowed the creation of a Requirement Analysis that is used as a metric to assess the publications that were recognised to be a part of the Automated Cyber Defence (ACD) literature. Through an extensive review of existing literature within automated blue and red teaming algorithms within custom ACO Gyms, it was discovered that Deep Reinforcement Learning (DRL) solutions were the most optimal algorithms for automated blue and red teaming as compared to Game Theoretic and Machine Learning solutions. This was primarily due to their ability to take sequential response for long-term and short-term goals. In addition, the review suggested the need for a parallel research and development of automated (blue and red) agents and ACO Gyms in order to accelerate research in both domains respectively. An extensive review was also conducted on existing open and closed-source gyms along with automated red and blue teaming implementations within them. The publications and implementations were assessed through the requirement analysis to find areas of further development within the literature. Through the requirement analysis of all publications of automated agents and ACO Gyms, specific challenges and gaps were discovered and elaborated. The challenges for the area of ACD included research within cybersecurity attack robustification of automated blue agents, realism of ACO Gyms, impact of incorrect actions and action/observation spaces within ACO Gyms. In contrast, the gaps included research within algorithmic attacks on the automated blue agent, explainable RL as automated blue agents and multi-agent automated blue agents. The aim of the challenges and gaps address the areas of future research and development within ACD in order for the transition of automated blue agents from ACO Gyms to networked systems deployed in the real world.

\bibliographystyle{ACM-Reference-Format}
\bibliography{acd-a-review}

\end{document}